\documentclass{elsart}
\usepackage{epsfig}
\usepackage{amssymb}

\begin{document}

\begin{frontmatter}

\title{
Cooperon propagator description of high temperature superconductivity
}

\author{C. Berthod and}
\author[ca]{B. Giovannini}
\ead{bernard.giovannini@physics.unige.ch}
\corauth[ca]{Corresponding author. Fax: +41-22-7026869}

\address{
DPMC, Universit\'e de Gen\`eve, 24 Quai Ernest-Ansermet,
1211 Gen\`eve 4, Switzerland
}

\begin{abstract}

A phenomenological description of the high-$T_c$ superconductors based on the
Cooperon propagator is presented. This model allows one to study the effects of
local pairing correlations and long-range phase fluctuations on the same
footing, both above and below $T_c$. Based on numerical calculations, it is
shown that the two types of correlations contribute to the gap/pseudogap in the
single-particle excitation spectra. The concourse of these two effects can
induce low energy states, which should be observable in underdoped materials at
very low temperature.

\end{abstract}


\end{frontmatter}

\small

The nature of the pseudogap phase in the high-$T_c$ cuprate superconductors
(HTS) is still the subject of lively debate and its elucidation may be one of
the keys to the understanding of these materials. At present we may perhaps,
from the experimental point of view, state the following: (1) The pseudogap ---
defined as a depletion of the single particle density of states (DOS) around
the Fermi energy, which appears gradually at a temperature $T^*$ --- goes
smoothly into the superconducting gap at $T_c$, as seen in STM\cite{19980105}
and ARPES\cite{19980503} measurements. (2) The STM measurements of vortices
below $T_c$ show that the DOS in the center of the vortex core ressembles
strongly the DOS observed above $T_c$ in the pseudogap
regime\cite{19980420,20000814}. (3) The fluctuations above $T_c$ (and outside
the 3D critical regime near $T_c$) seem to correspond to a phase fluctuation
regime, quantitatively well described by the Kosterlitz-Thouless (KT) theory
above $T_{\rm KT}$\cite{19990318,20000804}. (4) This phase fluctuation regime
seems to disappear at a characteristic temperature well below
$T^*$\cite{19990318,20000804}. One is therefore in need of a theoretical
framework which ({\it i\/}) is able to show a smooth transition of the DOS
across $T_c$, ({\it ii\/}) connects the phase fluctuation regime with the
properties of the KT theory, and ({\it iii\/}) relates the properties within a
vortex core to the properties of the pseudo-gapped phase. Such a theoretical
framework has been developed 40 years ago by Kadanoff and Martin
(KM)\cite{19611101}, who presented a theory of superconductivity entirely based
on the pair correlation function (Cooperon propagator). In a recent
paper\cite{phfluct}, we applied this framework for the $s$-wave symmetry. The
purpose of this communication is to extend these results to the $d$-wave
symmetry case.

In the KM formalism, the relation between the two-body Cooperon propagator
$
L({\bf r},\,{\bf r}';\,{\bf s}',\,{\bf s};\,\tau) =
\langle T_{\tau}\{
\psi_{\uparrow}^{ }({\bf r},\tau)
\psi_{\downarrow}^{ }({\bf r}',\tau)
\psi_{\downarrow}^{\dagger}({\bf s},0)
\psi_{\uparrow}^{\dagger}({\bf s}',0)
\} \rangle
$
and the self-energy is
	\begin{equation}\label{eq:Sigma}
		\Sigma({\bf r},\,{\bf s},\,\tau) = -\int d{\bf r}'d{\bf s}'\,
		V({\bf r},\,{\bf r}')
		L({\bf r},\,{\bf r}';\,{\bf s}',\,{\bf s};\,\tau)
		V({\bf s}',\,{\bf s})\,
		G_0({\bf s}',\,{\bf r}',\,-\tau),
	\end{equation}
where $G_0$ is the free Green's function and $V({\bf r},\,{\bf r}')$ is the
effective interaction between the electrons. In a translationally invariant
system with a short-ranged interaction, we may approximate the product $VLV$ in
Eq.~(\ref{eq:Sigma}) by the form
$
\varphi({\bf r}-{\bf r}')\,\varphi({\bf s}'-{\bf s})\,
\Lambda({\bf r}'-{\bf s}',\,\tau)
$,
where $\varphi({\bf r})$ reflects the symmetry of the short-range correlations
and $\Lambda({\bf r},\,\tau)$ describes their strength and long-range
properties. Thus Eq.~(\ref{eq:Sigma}) can be rewritten in Fourier-Matsubara
representation as:
	\begin{equation}\label{eq:Sigmak}
		\Sigma({\bf k},\,\omega_n) = k_{\rm B}T\sum_m\int\frac{d{\bf q}}{(2\pi)^2}
		\frac{\varphi^2({\bf k})\,\Lambda({\bf k}+{\bf q},\,\omega_n+\omega_m)}
		{-i\omega_m + \varepsilon({\bf q})},
	\end{equation}
where $\varepsilon({\bf k})$ is the free dispersion. The self-energy recovers
the BCS form when $\Lambda$ is independent of distance and time, i.e.
$\Lambda\propto\delta({\bf k}+{\bf q})\,\delta_{\omega_n,-\omega_m}$ in
Eq.~(\ref{eq:Sigmak}). In this formalism, the superconducting order is
therefore related to the long-range properties of the function $L$, rather than
to the strength of the superconducting correlations.

Along the lines of Ref.~\cite{phfluct}, we use the following phenomenological
model for $\Lambda$:
	\begin{equation}\label{eq:Lambda}
		\Lambda({\bf r},\,\tau)=
		\Delta_0^2\,e^{-r/\varrho_0}+\Delta_1^2\,e^{-r/\xi(T)},
	\end{equation}
where static ($\tau$-independent) correlations are assumed for simplicity. The
first term corresponds to the local correlations of range $\varrho_0$ and the
second term describes the phase fluctuations regime above $T_c$ --- $\xi(T)$
is the correlation length of the KT theory --- and the phase coherence below
$T_c$. Both types of correlations are supposed to have the same symmetry. With
this model, the self-energy on the real-frequency axis becomes
$
\Sigma({\bf k},\,\omega) =\varphi^2({\bf k})\int\frac{d{\bf q}}{(2\pi)^2}\,
\Lambda({\bf k}+{\bf q})/\left[\omega+i0^+ + \varepsilon({\bf q})\right]
$.
In the following, we consider $\Delta_0$ and $\Delta_1$ as temperature
independent parameters, and we calculate numerically the spectral function
$
A({\bf k},\,\omega) = -\frac{1}{\pi}{\rm Im}\left[\omega+i0^+ -
\varepsilon({\bf k}) - \Sigma({\bf k},\,\omega)\right]^{-1}
$
for a uniform two-dimensional square lattice, a $d$-wave factor $\varphi({\bf
k}) = \frac{1}{2}(\cos k_x-\cos k_y)$, and a dispersion $\varepsilon({\bf k})$
representative of the Bi$_2$Sr$_2$CaCu$_2$O$_{8+\delta}$ (Bi2212) band
structure\cite{19950701}.

\begin{figure}[tb!]
\leavevmode\begin{center}\epsfxsize8cm\epsfbox{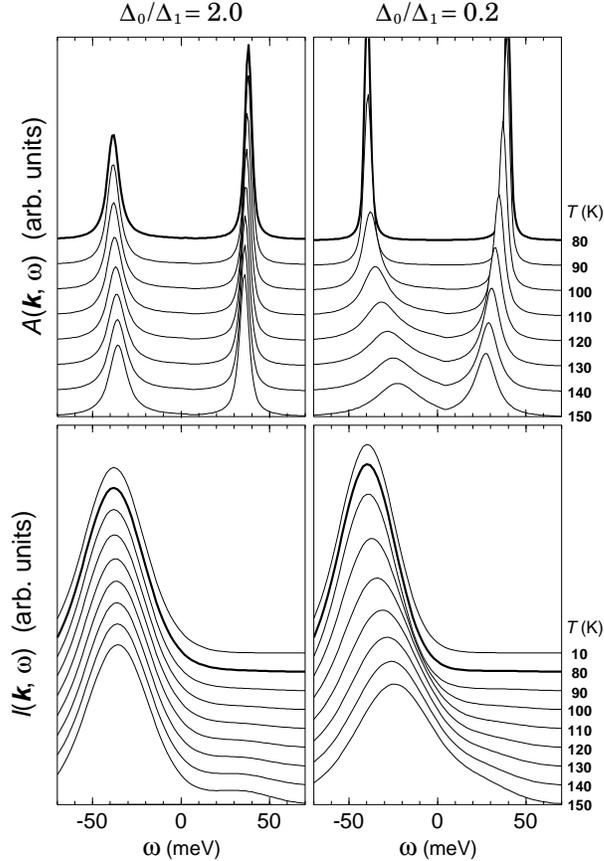}\end{center}
\caption{\label{fig:ARPES}
Upper panels: Spectral function at the Fermi point near $(\pi,\,0)$ for several
temperatures above $T_c$ in the strong (left) and weak (right) pseudogap cases.
The particle-hole asymmetry in the spectra is due to the underlying electronic
structure. Lower panels: Calculated ARPES intensity below and above $T_c$. The
model parameters are $T_c=80$~K, $\varrho_0=10\,a$, and
$[\Delta_0^2+\Delta_1^2]^{1/2}=40$~meV. The curves are offset vertically by
equal amounts. The thick curves correspond to $T=T_c$.
}
\end{figure}

The resulting $A({\bf k},\,\omega)$ at the antinodal point of the Fermi surface
is shown in Fig.~\ref{fig:ARPES} (upper panels) for several temperatures above
$T_c$. One can see marked differences between the strong ($\Delta_0>\Delta_1$)
and weak ($\Delta_0<\Delta_1$) pseudogap cases. In the former case, the
spectral function is dominated by the contribution of the local correlations
and remains basically temperature independent. In the latter case, the
quasiparticle peak broadens with increasing temperature, and moves toward the
Fermi energy. It is worth noting, in this connection, that the gap in the
spectral function does not relate solely to the {\em strength\/} of the
correlations --- or to the amplitude of the superconducting order parameter ---
but also to the {\em range\/} of these very correlations; this is particularly
clear in the right part of Fig.~\ref{fig:ARPES}, since $\Delta_1$ is the same
for all curves. Therefore, the spectral function alone does not provide a
reliable measure of the amplitude of the fluctuating order parameter above
$T_c$.

In order to compare our results with the experimental data, we must take into
account the angular ($\delta\theta\approx 5^{\circ}$) and energy
($\delta\omega\approx 15$~meV) resolutions of the ARPES measurements. We thus
estimate the ARPES intensity $I({\bf k},\,\omega)$ as a convolution of the
calculated $A({\bf k},\,\omega)\,f(\omega)$, $f$ the Fermi function, with a
Gaussian of appropriate width in ${\bf k}$ and $\omega$. The calculated $I({\bf
k},\,\omega)$ is shown in the lower panels of Fig.~\ref{fig:ARPES}. Because of
the ${\bf k}$-dependence of the spectral function, the effect of a finite
angular resolution is mainly to add a background in the ARPES signal. In the
strong pseudogap case, we find that the peak position is more or less
temperature independent, in agreement with the experimental results in
underdoped Bi2212\cite{19980503}. In the weak pseudogap case, however, the peak
approaches the Fermi level as observed in overdoped Bi2212\cite{19980503}. The
ARPES lineshapes in Fig.~\ref{fig:ARPES} do not show a sudden disappeareance of
the quasiparticle peak at $T_c$\cite{20000221}. This might be due to our
assumption of {\em static\/} correlations in Eq.~(\ref{eq:Lambda}). Short time
correlations above $T_c$ can be expected to decrease the quasiparticle lifetime
and eventually to induce a redistribution of the spectral weight over a large
energy window.

\begin{figure}[tb!]
\leavevmode\begin{center}\epsfxsize8cm\epsfbox{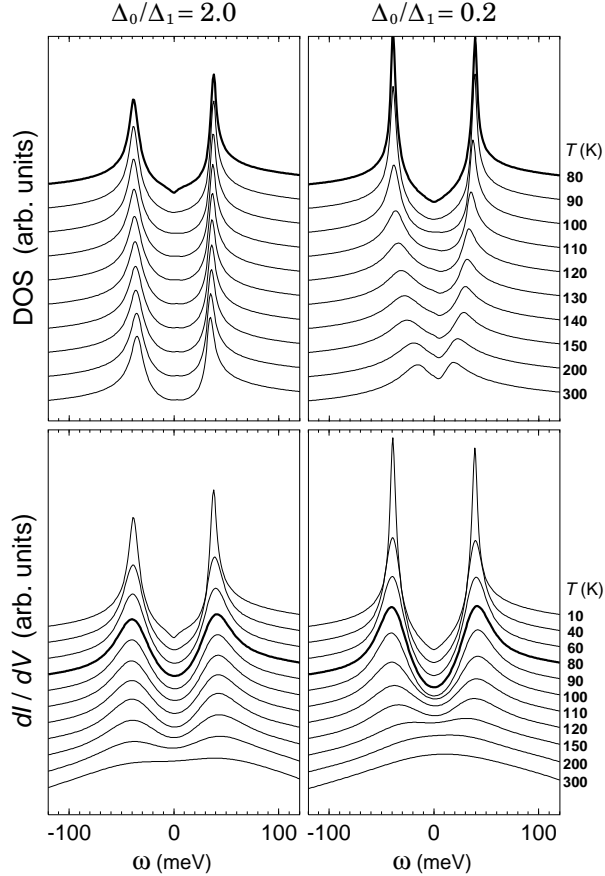}\end{center}
\caption{\label{fig:STM}
Upper panels: Density of states from $T_c$ to room temperature in the strong
(left) and weak (right) pseudogap cases. Lower panels: Calculated tunneling
conductance from $T=10$ to 300~K. The model parameters are the same as in
Fig.~\ref{fig:ARPES}. The curves are offset vertically by equal amounts.
The thick curves correspond to $T=T_c$.
}
\end{figure}

In Fig.~\ref{fig:STM} (upper panels), we plot the calculated density of states
(DOS) $N(\omega)=\int d{\bf k}\,A({\bf k},\,\omega)$ for temperatures above
$T_c$ (the DOS below $T_c$ is equal to the DOS at $T_c$ in the present model).
The lower panels show the tunneling conductance across $T_c$, calculated as
$dI/dV\propto\int d\varepsilon f'(\varepsilon-eV)\int d{\bf k}\,|T_{\bf k}|^2\,
A({\bf k},\,\varepsilon)$ with a uniform matrix element $T_{\bf k}\equiv T_0$.
The behavior of the DOS as a function of temperature parallels the behavior of
the spectral function in Fig.~\ref{fig:ARPES}. In the strong pseudogap case,
the effect of the local correlations is to broaden the DOS, resulting in
smaller coherence peaks and in a large zero-bias conductance. In addition, we
find characteristic shoulders near the Fermi energy, reminiscent of the
low-energy states observed in Bi2212 vortex cores\cite{19980420,20000814}. As
Fig.~\ref{fig:STM} shows, these states disappear above $T_c$, leading to a more
U-shaped spectrum. Moreover, we have verified that these structures are absent
if $\varrho_0\gtrsim 20\,a$, and that they do not appear if $\Delta_0=0$ or
$\Delta_1=0$. These findings indicate that the low energy states result from
the interplay of the short-range and long-range correlations. This is confirmed
by our numerical results in the weak pseudogap case. Indeed, subgap structures
can be seen in the DOS for $T>100$~K, i.e. a temperature range where $\xi(T)$
is smaller than $\varrho_0$.

In the tunneling spectra shown in Fig.~\ref{fig:STM} (lower panels), the
superconducting gap evolves smoothly across $T_c$ into a pseudogap, but the
coherence peaks and the gap structure is suppressed more rapidly in the weak
pseudogap case as the temperature is raised. Also, the width of the pseudogap
appears to be almost temperature independent in both strong and weak pseudogap
cases. These trends are consistent with the experimental observations in under-
and overdoped Bi2212\cite{19980105}, respectively.

In summary, based on a phenomenological theory of HTS connected to the
properties of the Cooperon propagator, we have investigated the effect of local
pairing correlations and long-range phase fluctuations on the spectral
properties of a 2-dimensional $d$-wave superconductor. The energy and
temperature dependence of the spectral functions was shown to vary with the
relative strength $\Delta_0/\Delta_1$ and the relative range $\varrho_0/\xi(T)$
of both types of correlations. In particular, low energy states were found to
appear in the DOS in two distinct regions of the parameter space:
$\Delta_0/\Delta_1>1$, $\varrho_0/\xi(T)\ll 1$ (strong pseudogap case at low
temperature) and $\Delta_0/\Delta_1<1$, $\varrho_0/\xi(T)>1$ (weak pseudogap
case at high temperature), i.e. when one type of correlation is stronger in
strength and weaker in range with respect to the other type of correlation.


\begin{thebibliography}{00}

\bibitem{19980105}
Ch. Renner, B. Revaz, J.-Y. Genoud, K. Kadowaki, and \O. Fischer,
Phys. Rev. Lett. {\bf 80}, 149 (1998)

\bibitem{19980503}
M. R. Norman, M. Randeria, H. Ding, and J. C. Campuzano,
Phys. Rev. B {\bf 57}, R11093 (1998).


\bibitem{19980420}
Ch. Renner, B. Revaz, K. Kadowaki, I. Maggio-Aprile, and \O. Fischer,
Phys. Rev. Lett. {\bf 80}, 3606 (1998);
B.W. Hoogenboom, Ch. Renner, B. Revaz, I. Maggio-Aprile, and \O. Fischer,
Physica C {\bf 332}, 440 (2000).

\bibitem{20000814} S. H. Pan, E. W. Hudson, A. K. Gupta, K.-W. Ng,
H. Eisaki, S. Uchida, and J. C. Davis,
Phys. Rev. Lett. {\bf 85}, 1536 (2000).

\bibitem{19990318}
J. Corson, R. Mallozzi, J. Orenstein, J. N. Eckstein, and I. Bozovic,
Nature {\bf 398}, 221 (1999).

\bibitem{20000804}
Z. A. Xu, N. P. Ong, Y. Wang, T. Kakeshita, and S. Uchida,
Nature {\bf 406}, 486 (2000).

\bibitem{19611101}
L. P. Kadanoff and P. C. Martin, Phys. Rev. {\bf 124}, 670 (1961).

\bibitem{phfluct}
B. Giovannini and C. Berthod, to appear in Phys. Rev. B.

\bibitem{19950701}
M. R. Norman, R. Randeria, H. Ding, and J. C. Campuzano,
Phys. Rev. B {\bf 52}, 615 (1995).
In order to distinguish the peaks induced by the pairing correlations from the
peak due to the van-Hove singularity, we shifted the chemical potential by
$-30$~meV so that it coincides with the singularity.

\bibitem{20000221}
A. Kaminski, J. Mesot, H. Fretwell, J. C. Campuzano, M. R. Norman, M. Randeria,
H. Ding, T. Sato, T. Takahashi, T. Mochiku, K. Kadowaki, and H. Hoechst,
Phys. Rev. Lett. {\bf 84}, 1788 (2000).



\end{thebibliography}
\end{document}